\documentstyle[12pt,epsf]{article}
\begin{document}
\baselineskip=24pt
\def\rd{{\rm d}}
\newcommand{\Lamb}{\Lambda_{b}}
\newcommand{\Lamc}{\Lambda_{c}}
\newcommand{\Lam}{\Lambda}
\newcommand{\dsp}{\displaystyle}
\newcommand{\nn}{\nonumber}
\newcommand{\ra}{\rightarrow}
\newcommand{\dfr}[2]{ \displaystyle\frac{#1}{#2} }
\newcommand{\Lag}{\Lambda \scriptscriptstyle _{ \rm GR} } 
\renewcommand{\baselinestretch}{1.5}
\begin{titlepage}
\vspace{-20ex}
\vspace{1cm}
\begin{flushright}
\vspace{-3.0ex} 
    {\sf ADP-99-26/T366} \\
\vspace{-2.0mm}
\vspace{5.0ex}
\end{flushright}

\centerline{\Large\sf Direct CP Violation in Charmed Hadron Decays via 
$\rho-\omega$ Mixing}
\vspace{6.4ex}
\centerline{\large\sf  	X.-H. Guo$^{1,2}$ and A.W. Thomas$^{1}$}
\vspace{3.5ex}
\centerline{\sf $^1$ Department of Physics and Mathematical Physics,}
\centerline{\sf and Special Research Center for the Subatomic Structure of
Matter,}
\centerline{\sf University of Adelaide, SA 5005, Australia}
\centerline{\sf $^2$ Institute of High Energy Physics, Academia Sinica,
Beijing 100039, China}
\centerline{\sf e-mail:  xhguo@physics.adelaide.edu.au,
athomas@physics.adelaide.edu.au}
\vspace{6ex}
\begin{center}
\begin{minipage}{5in}
\centerline{\large\sf 	Abstract}
\vspace{1.5ex}
\small {We study the possibility of obtaining large direct CP violation in the 
charmed hadron decays
$D^+ \ra \rho^+ \rho^0 (\omega) \ra \rho^+ \pi^+\pi^-$,   
$D^+ \ra \pi^+ \rho^0 (\omega) \ra \pi^+ \pi^+\pi^-$,
$D^0 \ra \phi  \rho^0 (\omega) \ra\phi \pi^+\pi^-$,  
$D^0 \ra \eta  \rho^0 (\omega) \ra\eta \pi^+\pi^-$,  
$D^0 \ra \eta'  \rho^0 (\omega) \ra\eta' \pi^+\pi^-$,  
$D^0 \ra \pi^0 \rho^0 (\omega) \ra \pi^0 \pi^+\pi^-$, and
$\Lambda_c \ra p \rho^0 (\omega) \ra p \pi^+\pi^-$ 
via $\rho-\omega$ mixing. 
The analysis is carried out in the factorization approach. The CP violation
parameter depends on the effective parameter, $N_{c}$,
which is relevant to hadronization dynamics of each decay channel and
should be determined by experiments. 
It is found that for fixed $N_{c}$
the CP violation parameter reaches its maximum value when the invariant
mass of the $\pi^+\pi^-$ pair is in the vicinity of the $\omega$ resonance.
For $D^0 \ra \pi^0 \rho^0 (\omega) \ra \pi^0 \pi^+\pi^-$ and
$\Lambda_c \ra p \rho^0 (\omega) \ra p \pi^+\pi^-$ 
the maximum CP violating asymmetry is bigger than $1\%$ 
when $1.98\leq N_c \leq 1.99$ and $1.95\leq N_c \leq 2.02$, respectively. 
We also estimate the decay branching ratios for $D^0 \ra \pi^0 \rho^0$
and $\Lambda_c \ra p \rho^0$ for these values of $N_c$, which should be tested
by future experimental data.
}

\end{minipage}
\end{center}

\vspace{0.5cm}

\end{titlepage}
\vspace{0.2in}
{\large\bf I. Introduction}
\vspace{0.2in}

Although CP violation has been known in the neutral kaon system for more than 
three decades its dynamical origin still remains an open
problem. Besides the kaon system, the study of CP violation in heavy quark 
systems has been a subject of intense interest and is important in 
understanding whether the Standard Model provides a
correct description of this phenomenon through 
the Cabbibo-Kobayashi-Maskawa (CKM) matrix. Actually there have been many
theoretical studies in the area of CP violation in beauty and charm systems
and some experimental projects have been proposed\cite{kaplan}. 

Recent studies of direct CP violation in the $B$ meson system\cite{carter}
have suggested that large CP-violating asymmetries should  be observed in 
forthcoming experiments. However, in the charm sector, the CP violation
is usually predicted to be small. A rough estimate of CP violation in  
charmed systems gives an asymmetry parameter which is typically smaller than
$10^{-3}$ due to the suppression of the CKM matrix elements\cite{burdman}.  
By introducing large final-state-interaction phases provided by nearby 
resonances, Buccella {\it et al.} predicted larger CP violation, namely a 
few times $10^{-3}$\cite{buccella}. On the other hand, experimental 
measurements in some decay channels are consistent with zero 
asymmetry\cite{frabetti}. 

Direct CP violation occurs through the interference of
two amplitudes with different weak and strong phases. The
weak phase difference is determined by the CKM matrix elements and
the strong phase is usually very uncertain. 
In Refs.\cite{eno, tony}, the authors studied direct CP violation 
in hadronic $B$ decays 
through the interference of tree and penguin diagrams, where
$\rho-\omega$ mixing was used to obtain a large strong
phase (as required for large CP violation). This mechanism was also
applied to the hadronic decays of the heavy baryon, $\Lambda_b$, 
where even larger CP violation may be possible\cite{ga}. In the present paper
we will investigate direct CP violation
in the hadronic decays of charmed hadrons, involving the same mechanism, with
the aim of finding channels which may exhibit large CP asymmetry.

Since we are considering direct CP violation, we have to consider
hadronic matrix elements for both tree and penguin diagrams which are 
controlled by the effects of nonperturbative QCD and hence are uncertain.
In our discussions we will use the factorization approximation
so that one of the currents in the nonleptonic decay Hamiltonian is 
factorized out and generates a meson. Thus the decay amplitude of the 
two body nonleptonic decay
becomes the product of two matrix elements, one related to the decay
constant of the factorized meson and the other to the weak transition
matrix element between two hadrons. There have been some discussions
of the plausibility of factorization\cite{bjorken}\cite {dugan},
and this approach may be a good approximation in energetic decays. 
In some recent work corrections to the factorization approximation have 
also been considered by 
introducing some phenomenological nonfactorizable parameters which
depend on the specific decay channels and should be determined by experimental
data\cite{chengn, neubert, ali, buras1}.  

The effective Hamiltonian for the $\Delta S=1$, weak, nonleptonic decays has 
been discussed in detail in Refs.\cite{buras2}\cite{ciuchini}, where the
Wilson coefficients for the tree and penguin operators were
obtained to the next-to-leading order QCD and QED corrections by  
calculating the $10\times 10$, two-loop, anomalous dimension matrix.
The dependence of the Wilson coefficients on renormalization scheme,  
gauge and infra-red cutoff was also discussed. The formalism can be extended
to the charmed hadron nonleptonic decays in a straightforward way.

The remainder of this paper is organized as follows. In Section II we
calculate the six Wilson coefficients of tree and QCD penguin
operators to the next-to-leading order QCD corrections by applying the 
results of Refs.\cite{buras2}\cite{ciuchini}. Then in Section III we 
give the formalism for the CP-violating asymmetry in charmed
hadron nonleptonic decays and show numerical results. 
Finally, Section IV is reserved for a summary and some discussion.

\vspace{0.2in}
{\bf II. The effective Hamiltonian for nonleptonic charmed hadron decays} 
\vspace{0.2in}

In order to calculate direct CP violation in nonleptonic, charmed hadron decays
we use the following effective weak Hamiltonian, which is Cabibbo 
first-forbidden, based on the operator product expansion,
\begin{equation}
{\cal H}_{\Delta C=1} = {G_F\over \sqrt{2}}[\sum_{q=d,s}
V_{uq}V^*_{cq}(c_1 O^q_1 + c_2 O^q_2)
 - V_{ub}V^*_{cb}\sum^{6}_{i=3} c_i O_i] + H.C..\;
\label{2a}
\vspace{2mm}
\end{equation}
Here $c_i\;(i=1,...,6)$ are the Wilson coefficients and
the operators $O_i$ have the following expressions:   
\begin{eqnarray}
O^{q}_1&=& \bar u_\alpha \gamma_\mu(1-\gamma_5)q_\beta\bar
q_\beta\gamma^\mu(1-\gamma_5)c_\alpha,\;\;
O^{q}_2= \bar u \gamma_\mu(1-\gamma_5)q\bar
q\gamma^\mu(1-\gamma_5)c,\nn\\
O_3&=& \bar u \gamma_\mu(1-\gamma_5)c \sum_{q'}
\bar q' \gamma^\mu(1-\gamma_5) q',\;\;
O_4 = \bar u_\alpha \gamma_\mu(1-\gamma_5)c_\beta \sum_{q'}
\bar q'_\beta \gamma^\mu(1-\gamma_5) q'_\alpha,\nn\\
O_5&=&\bar u \gamma_\mu(1-\gamma_5)c \sum_{q'} \bar q'
\gamma^\mu(1+\gamma_5)q',\;\;
O_6 = \bar u_\alpha \gamma_\mu(1-\gamma_5)c_\beta \sum_{q'}
\bar q'_\beta \gamma^\mu(1+\gamma_5) q'_\alpha,\nn\\
&&
\label{2b}
\vspace{2mm}
\end{eqnarray}
where $\alpha$ and $\beta$ are color indices, and 
$q'=u,\;d,\;s$. In Eq.(\ref{2b})
$O_1$ and $O_2$ are the tree operators, while $O_{3}-O_{6}$ are QCD 
penguin operators. In the Hamiltonian we have omitted the operators 
associated with electroweak penguin diagrams.

The Wilson coefficients, $c_i\;(i=1,...,6)$, are calculable
in perturbation theory by using the renormalization group.
The solution has the following form,
\begin{equation}
{\bf C}(\mu) = U (\mu, m_W){\bf C}(m_W),
\label{2c}
\vspace{2mm}
\end{equation}
where $U (\mu, m_W)$ describes the QCD evolution which sums the logarithms
$(\alpha_s {\rm ln}(m_W^2/\mu^2))^n$ (leading-log approximation) and 
$\alpha_s(\alpha_s {\rm ln}(m_W^2/\mu^2))^n$ (next-to-leading order).
In Refs.\cite{buras2} \cite{ciuchini} it was shown that $U (m_1, m_2)$
can be written as
\begin{equation}
U (m_1, m_2)=\left(1+\frac{\alpha_s(m_1)}{4\pi} J\right)U^0 (m_1, m_2)
\left(1-\frac{\alpha_s(m_2)}{4\pi}J\right),
\label{2d}
\vspace{2mm}
\end{equation}
where $U^0 (m_1, m_2)$ is the evolution matrix in the leading-log 
approximation and the matrix $J$ summarizes the next-to-leading 
order corrections to this evolution.

The evolution matrices $U^0 (m_1, m_2)$ and $J$ can be obtained by calculating
the appropriate one- and two-loop diagrams respectively. The initial 
conditions, ${\bf C}(m_W)$, 
are determined by matching the full theory and the effective
theory at the scale $m_W$. At the scale $m_c$ the Wilson coefficients
are given by
\begin{equation}
{\bf C}(m_c) = U_4 (m_c, m_b) M(m_b) U_5 (m_b, m_W){\bf C}(m_W),
\label{2e}
\vspace{2mm}
\end{equation}
where $U_f (m_1, m_2)$ is the evolution matrix from $m_2$ to $m_1$ with
$f$ active flavors and $M(m_b)$ is the quark-threshold matching matrix at
$m_b$. Since the strong interaction is independent of quark flavors, the
matrices $U_4 (m_c, m_b)$, $ U_5 (m_b, m_W)$ and $ M(m_b)$ are the same
as those in b-decays. Hence, using the expressions for $U^0 (m_1, m_2)$,
$J$ and $M(m_b)$ given in \cite{buras2}\cite{ciuchini}, we can obtain 
${\bf C}(m_c)$.

In general, the Wilson coefficients depend on the renormalization scheme.
The scheme-independent Wilson coefficients $\bar{\bf C}(\mu)$ are introduced
by the following equation
\begin{equation}
\bar{\bf C}(\mu) = \left( 1+ \frac{\alpha_s}{4\pi}R^T \right){\bf C}(\mu),
\label{2f}
\vspace{2mm}
\end{equation}
where $R$ is the renormalization matrix associated with 
the four-quark operators 
$O_i (i=1,...,6)$ in Eq.(\ref{2b}), at the scale $m_W$. The scheme-independent
Wilson coefficients have been used in the
literature\cite{buccella, he, fleischer}.
However, since $R$ depends on the infra-red regulator\cite{buras2}, 
$\bar{\bf C}(\mu)$ also carries such a dependence. In the present paper we 
have chosen to use the scheme-independent Wilson coefficients.

From Eqs.(\ref{2e})(\ref{2f}) and the expressions for the matrices 
$U_f(m_1,m_2)$, $M(m_b)$ and $R$ in \cite{buras2}\cite{ciuchini} we obtain the
following scheme-independent Wilson coefficients for c-decays at the
scale $m_c=1.35$GeV:
\begin{eqnarray}
\bar c_1 &=& -0.6941, \;\;\bar c_2 = 1.3777, \;\;\bar c_3 = 0.0652, \nn\\
\bar c_4 &=& -0.0627,\;\;\bar c_5 = 0.0206,\;\; \bar c_6 = -0.1355.
\label{2g}
\vspace{2mm}
\end{eqnarray}
In obtaining Eq.(\ref{2g}) we have taken $\alpha_s (m_Z) =0.118$ which leads to
$\Lambda_{\rm QCD}^{(5)}=0.226$GeV and $\Lambda_{\rm QCD}^{(4)}=0.329$GeV.
To be consistent, the matrix elements of the operators
$O_i$ should also be renormalized to the one-loop order
since we are working to the next-to-leading order for the 
Wilson coefficients. This results in 
effective Wilson coefficients, $c'_i$, which satisfy the constraint
\begin{equation}
c_i (m_c) \langle O_i (m_c) \rangle =c'_i \langle O_{i} \rangle^{\rm tree},
\label{2h}
\vspace{2mm}
\end{equation}
where $\langle O_i (m_c) \rangle$ are the matrix elements, renormalized to 
the one-loop order.  The relations between $c'_i$ and $c_i$ 
read\cite{he, fleischer}
\begin{eqnarray}
c'_1 &=& \bar c_1,\;\; c'_2 = \bar c_2, \;\;c'_3 = \bar c_3 - P_s/3, \nn\\
c'_4 &=& \bar c_4 +P_s,\;\;c'_5 =\bar c_5 - P_s/3,\;\;c'_6 =\bar c_6 + P_s,
\label{2i}
\vspace{2mm}
\end{eqnarray}
where 
$$P_s = (\alpha_s(m_c)/8\pi) (10/9 +G(m,m_c,q^2))\bar c_2,$$
with 
$$G(m,m_c,q^2) = 4\int^1_0 {\rm d}x x(1-x) {\rm ln}{m^{2}-x(1-x)q^2\over
m_c^2}.$$
Here $q^2$ is the momentum transfer of the gluon in the penguin
diagram and $m$ is the mass of the quark in the loop of the penguin diagram
\footnote{$m$ could be $m_d$ or $m_s$. However, the numerical values
of $c'_i$ change by at most 4\% when we change $m$ from $m_d$ to $m_s$.
Therefore, we ignore this difference in our calculations, setting $m=m_s$.}. 
$G(m,m_c,q^2)$ has the following explicit expression \cite{kramer}
\begin{eqnarray}
{\rm Re}G&=&\frac{2}{3}\left({\rm ln}\frac{m^2}{m_c^2}-\frac{5}{3}
-4\frac{m^2}{q^2}+(1+2\frac{m^2}{q^2})\sqrt{1-4\frac{m^2}{q^2}}
{\rm ln}\frac{1+\sqrt{1-4\frac{m^2}{q^2}}}{1-\sqrt{1-4\frac{m^2}{q^2}}}
\right),\nn\\
{\rm Im}G&=&-\frac{2}{3}\pi\left(1+2\frac{m^2}{q^2}\right)
\sqrt{1-4\frac{m^2}{q^2}}.
\label{2j}
\vspace{2mm}
\end{eqnarray}

Based on simple arguments at the quark level, the value
of $q^2$ is chosen in the range $0.3 < q^2/m_{c}^2 < 0.5$\cite{eno, tony}. 
From Eqs.(\ref{2g}), (\ref{2i}) and (\ref{2j}) we can obtain 
numerical values of $c'_i$. When $q^2/m_{c}^2=0.3$,
\begin{eqnarray}
c'_1 &=&-0.6941, \;\; c'_2=1.3777,\nn\\
c'_3 &=& 0.07226+0.01472i,\;\; 
c'_4 = -0.08388-0.04417i,\nn\\ 
c'_5 &=& 0.02766+0.01472i,\;\; 
c'_6 = -0.1567-0.04417i,
\label{2k}
\vspace{2mm}
\end{eqnarray}
and when $q^2/m_{c}^2=0.5$,
\begin{eqnarray}
c'_1 &=&-0.6941, \;\; c'_2=1.3777,\nn\\
c'_3 &=& 0.06926+0.01483i,\;\; 
c'_4 = -0.07488-0.04448i,\nn\\ 
c'_5 &=& 0.02466+0.01483i,\;\; 
c'_6 = -0.1477-0.04448i.
\label{2l}
\vspace{2mm}
\end{eqnarray}

In calculating the matrix elements of the Hamiltonian (\ref{2a}), we can then
simply use the effective Wilson coefficients in 
Eqs.(\ref{2k})({\ref{2l}) to multiply
the tree-level matrix elements of the operators $O_i (i=1,...,6)$.

\vspace{0.2in}
{\large\bf III. CP violation in charmed hadron decays}
\vspace{0.2in}

{\bf III.1 Formalism for CP violation in charmed hadron decays}
\vspace{0.2in}

The formalism for CP violation in $B$ and $\Lamb$ hadronic decays 
\cite{eno, tony, ga} can be generalized to the case of charmed hadrons
in a straightforward manner. Let $H_c$ denote a charmed hadron which
could be $D^{\pm}$, $D^0$, or $\Lambda_c$. 
The amplitude, $A$, for the decay $H_c \rightarrow f \pi^+ \pi^-$ 
($f$ is a decay product) is:
\begin{equation}
A = \langle \pi^+\pi^- f | {\cal H}^{\rm T} | H_c \rangle
+ \langle \pi^+\pi^- f | {\cal H}^{\rm P} | H_c \rangle,
\label{3a}
\vspace{2mm}
\end{equation}
where ${\cal H}^{\rm T}$ and ${\cal H}^{\rm P}$ are the 
Hamiltonians for the tree and penguin operators, respectively. 

The relative magnitude and phases of these two diagrams are defined 
as follows:
\begin{eqnarray}
A &=& \langle \pi^+\pi^- f | {\cal H}^{\rm T} | H_c \rangle \left[
1 + re^{i\delta} e^{i\phi} \right], \nn\\
\bar{A} &=& \langle \pi^+\pi^- \bar{f} | {\cal H}^{\rm T} | \bar H_c
\rangle \left[1 + re^{i\delta} e^{-i\phi} \right],
\label{3b}
\vspace{2mm}
\end{eqnarray}
where $\delta$ and $\phi$ are strong and weak phases, respectively.
$\phi$ arises from the CP-violating phase in the CKM matrix, and 
it is arg$[V_{ub}V^{*}_{cb}/(V_{uq}V^{*}_{cq})]$ for the $c\rightarrow q$
transition ($q=d$ or $s$). The parameter $r$ is defined as
\begin{equation}
r \equiv \left| \frac{\langle \pi^+\pi^- f | {\cal H}^{\rm P} | 
H_c \rangle}
{\langle \pi^+\pi^- f | {\cal H}^{\rm T} | H_c \rangle}\right|.
\label{3c}
\vspace{2mm}
\end{equation}

The CP-violating asymmetry, $a$, can be written as: 
\begin{equation}
a \equiv { | A |^2 - |{\overline A}|^2 
\over | A |^2 + |{\overline A}|^2 }
= {-2r \sin \delta \sin \phi 
\over 1 + 2r\cos \delta \cos \phi + r^2 }.
\label{3d}
\vspace{2mm}
\end{equation}
It can be seen from Eq.(\ref{3d}) that both weak and strong
phases are needed to produce CP violation. Since in $r$ there is strong
suppression from the ratio of the CKM matrix elements,
$[V_{ub}V^{*}_{cb}/(V_{uq}V^{*}_{cq})]$, which is of the order $10^{-3}$
\cite{burdman} (for both $q=d$ and $q=s$ this suppression is 
$0.62\times 10^{-3}$, see Eqs.(\ref{3p}) and (\ref{3u}) in III.2), 
usually the CP violation in charmed hadron decays is predicted to be small.

The weak phase $\phi$ for a specific physical process is fixed.
In order to obtain possible large CP violation, we need some mechanism
to produce either large $\sin \delta$ or large $r$.  
$\rho-\omega$ mixing has the dual advantages that the strong phase difference
is large (passing through $90^\circ$ at the $\omega$ resonance) and well known.
In this scenario one has \cite{tony, ga}
\begin{eqnarray}
\langle \pi^+\pi^- f | {\cal H}^{\rm T} | H_c \rangle 
&=& {g_{\rho} \over s_\rho s_\omega} \tilde\Pi_{\rho\omega} t_{\omega}
  + { g_{\rho} \over s_\rho } t_\rho, \label{3e1}\\
\langle \pi^+\pi^- f | {\cal H}^{\rm P} | H_c \rangle 
&=& {g_{\rho} \over s_\rho s_\omega} \tilde\Pi_{\rho\omega} p_{\omega}
  + { g_{\rho} \over s_\rho } p_\rho, 
\label{3e2}
\vspace{2mm}
\end{eqnarray}
where $t_{\rm V}$ (V=$\rho$ or $\omega$) is the tree 
and $p_{\rm V}$ is the penguin amplitude for 
producing a vector meson, ${\rm V}$, by $H_c \rightarrow f {\rm V}$;
$g_\rho$ is the coupling for $\rho^0 \ra \pi^+\pi^-$; 
$\tilde\Pi_{\rho\omega}$ is the effective  $\rho - \omega$ mixing amplitude
and $s_{\rm V}^{-1}$ is the propagator of V,
$s_{\rm V}=s - m_{\rm V}^2 + i m_{\rm V} \Gamma_{\rm V}$,
with $\sqrt{s}$ being the invariant mass of the $\pi^+ \pi^-$ pair.
The numerical values 
for the $\rho-\omega$ mixing parameter are\cite{tony, pionff97, pionffn97}:
${\rm Re}\tilde\Pi_{\rho\omega}(m_\omega^2) = -3500\pm 300
{\rm MeV}^2,\;\; {\rm Im}\tilde\Pi_{\rho\omega}(m_\omega^2)= -300 
\pm 300 {\rm MeV}^2.$ The direct coupling $\omega \ra \pi^+\pi^-$ is 
effectively absorbed into $\tilde\Pi_{\rho\omega}$\cite{pionffn97}.

Defining
\begin{equation}
{p_\omega \over t_\rho} \equiv r' e^{i(\delta_q + \phi)}, \quad
{t_\omega \over t_\rho} \equiv \alpha e^{i \delta_\alpha}, \quad
{p_\rho \over p_\omega} \equiv \beta e^{i \delta_\beta},
\label{3f}
\vspace{2mm}
\end{equation}
where $\delta_\alpha$, $\delta_\beta$ and $\delta_q$ are strong phases, one
has the following expression for $r$ and $\delta$,
\begin{equation}
re^{i\delta} = r' e^{i\delta_q} \frac{
 \tilde\Pi_{\rho\omega} + \beta e^{i\delta_\beta}s_\omega}{s_\omega
+\tilde\Pi_{\rho\omega} \alpha e^{i\delta_\alpha}}.
\label{3g}
\vspace{2mm}
\end{equation}

It will be shown that in the factorization approach, for all the decay 
processes $H_c \ra f \pi^+\pi^-$ we are considering, 
$\alpha e^{i\delta_\alpha}$ is real (see III.2 for details). Therefore, we let
\begin{equation}
\alpha e^{i\delta_\alpha}=g,
\label{3h}
\vspace{2mm}
\end{equation}
where $g$ is a real parameter.
Letting
\begin{equation}
\beta e^{i\delta_\beta}=b+ci,\;\;r' e^{i\delta_q}=d+ei, 
\label{3i}
\vspace{2mm}
\end{equation}
and using Eq.(\ref{3g}), we obtain the following result when $\sqrt{s}\sim
m_\omega$,
\begin{equation}
re^{i\delta} = \frac{C+Di}{(s-m_{\omega}^{2}+g 
{\rm Re}\tilde\Pi_{\rho\omega})^2
+(g {\rm Im}\tilde\Pi_{\rho\omega}+m_\omega \Gamma_\omega)^2},
\label{3j}
\vspace{2mm}
\end{equation}
where
\begin{eqnarray}
C&=&(s-m_{\omega}^{2}+g {\rm Re}\tilde\Pi_{\rho\omega})\{d
[{\rm Re}\tilde\Pi_{\rho\omega}+b(s-m_{\omega}^{2})-cm_\omega \Gamma_\omega]
\nn\\
&&
-e[{\rm Im}\tilde\Pi_{\rho\omega}+bm_\omega \Gamma_\omega+c(s-m_{\omega}^{2})]
\} \nn\\
& &+(g {\rm Im}\tilde\Pi_{\rho\omega}+m_\omega \Gamma_\omega)
\{e[{\rm Re}\tilde\Pi_{\rho\omega}+b(s-m_{\omega}^{2})-cm_\omega \Gamma_\omega]
\nn\\
&&+d[{\rm Im}\tilde\Pi_{\rho\omega}+bm_\omega \Gamma_\omega+c(s-m_{\omega}^{2})]
\}, \nn\\
D&=&(s-m_{\omega}^{2}+g {\rm Re}\tilde\Pi_{\rho\omega})\{e
[{\rm Re}\tilde\Pi_{\rho\omega}+b(s-m_{\omega}^{2})-cm_\omega \Gamma_\omega]
\nn\\
&&+d[{\rm Im}\tilde\Pi_{\rho\omega}+bm_\omega \Gamma_\omega+c(s-m_{\omega}^{2})]
\} \nn\\
& &-(g {\rm Im}\tilde\Pi_{\rho\omega}+m_\omega \Gamma_\omega)
\{d[{\rm Re}\tilde\Pi_{\rho\omega}+b(s-m_{\omega}^{2})-cm_\omega \Gamma_\omega]
\nn\\
&&
-e[{\rm Im}\tilde\Pi_{\rho\omega}+bm_\omega \Gamma_\omega+c(s-m_{\omega}^{2})]
\}. 
\label{3k}
\vspace{2mm}
\end{eqnarray}

The weak phase comes from $[V_{ub}V^{*}_{cb}/(V_{uq}V^{*}_{cq})]$. If the 
operators $O_1^d, O_2^d$ contribute to the decay processes we have
\begin{eqnarray}
{\rm sin}\phi |_d&=&\frac{\eta}{\sqrt{[\rho +A^2\lambda^4 (\rho^2+\eta^2)]^2
+\eta^2}}, \nn\\
{\rm cos}\phi |_d&=&-\frac{\rho +A^2\lambda^4 (\rho^2+\eta^2)}
{\sqrt{[\rho +A^2\lambda^4 (\rho^2+\eta^2)]^2
+\eta^2}}, 
\label{3l1}
\vspace{2mm}
\end{eqnarray}
while if $O_1^s$ and $O_2^s$ contribute, we have
\begin{eqnarray}
{\rm sin}\phi |_s&=&-\frac{\eta}{\sqrt{\rho^2+\eta^2}}, \nn\\
{\rm cos}\phi |_s&=&\frac{\rho}
{\sqrt{\rho^2+\eta^2}}, 
\label{3l2}
\vspace{2mm}
\end{eqnarray}
where we have used the Wolfenstein parametrization \cite{wolf} for the
CKM matrix elements. In order to obtain $r{\rm sin}\delta$, 
$r{\rm cos}\delta$ and $r$ we need to calculate 
$\beta e^{i\delta_\beta}$ and $r' e^{i\delta_q}$. This will be done in the
next subsection.

\vspace{0.2in}
{\bf III.2 CP violation in $H_c\rightarrow f \pi^+\pi^-$} 
\vspace{0.2in}

In the following we will calculate the CP-violating asymmetries 
in $H_c\rightarrow f\pi^+\pi^-$.  In the factorization approximation
$\rho^0 (\omega)$ is generated by one current which has the proper
quantum numbers in the Hamiltonian in Eq.(\ref{2a}). In the following
we will consider the decay processes 
$D^+ \ra \rho^+ \rho^0 (\omega) \ra \rho^+ \pi^+\pi^-$,   
$D^+ \ra \pi^+ \rho^0 (\omega) \ra \pi^+ \pi^+\pi^-$,
$D^0 \ra \phi  \rho^0 (\omega) \ra\phi \pi^+\pi^-$,  
$D^0 \ra \eta  \rho^0 (\omega) \ra\eta \pi^+\pi^-$,  
$D^0 \ra \eta'  \rho^0 (\omega) \ra\eta' \pi^+\pi^-$,  
$D^0 \ra \pi^0 \rho^0 (\omega) \ra \pi^0 \pi^+\pi^-$, and
$\Lambda_c \ra p \rho^0 (\omega) \ra p \pi^+\pi^-$, 
individually.

First we consider $D^+\rightarrow \rho^+ \rho^0 (\omega)$. After factorization,
the contribution to $t^{\rho^+}_{\rho}$ 
(the superscript denotes the decay product $f$ in $H_c \ra f \pi^+
\pi^-$) from the tree level operator $O^d_1$ is
\begin{equation}
\langle \rho^+\rho^0  |O^d_1 |D^+ \rangle = \langle \rho^0 |(\bar{d}d)|0\rangle
\langle \rho^+ | (\bar{u} c) |D^+ \rangle \equiv T_1,
\label{3m}
\vspace{2mm}
\end{equation}
where $(\bar{d}d)$ and $(\bar{u} c)$ denote the V-A currents. 
If we ignore isospin violating effects, then the matrix element of 
$O^d_2$ is the same as that of $O^d_1$. After adding the contributions
from Fierz transformation of $O^d_1$ and $O^d_2$ we have
\begin{equation}
t^{\rho^+}_{\rho}=(c'_1+c'_2) \left(1+1/N_c\right)T_1,
\label{3n}
\vspace{2mm}
\end{equation}
where we have omitted the CKM matrix elements in the expression of 
$t^{\rho^+}_{\rho}$.
Since in Eq.(\ref{3n}) we have neglected the color-octet 
contribution, which is nonfactorizable and difficult to calculate,
$N_c$ should be treated as an effective parameter which depends on 
the hadronization dynamics of different decay channels. 
In the same way we find that $t^{\rho^+}_{\omega} = -t^{\rho^+}_{\rho}$, 
so that, from Eq.(\ref{3f}), we have
\begin{equation}
(\alpha e^{i\delta_\alpha})^{\rho^+}=-1.
\label{3o1}
\vspace{2mm}
\end{equation}

The penguin operator contributions, $p^{\rho^+}_{\rho}$ and 
$p^{\rho^+}_{\omega}$, can be evaluated in the same way with the aid of the 
Fierz identities.
From  Eq.(\ref{3f}) we have
\begin{equation}
(\beta e^{i\delta_\beta})^{\rho^+}=0, 
\label{3o2}\\
\vspace{2mm}
\end{equation}
and
\begin{equation}
(r' e^{i\delta_q})^{\rho^+}=2\frac{(c'_3+c'_4)(1+\frac{1}{N_c})
+c'_5+\frac{1}{N_c}c'_6}{(c'_1+c'_2)(1+\frac{1}{N_c})}
\left|\frac{V_{ub}V^{*}_{cb}}{V_{ud}V^{*}_{cd}}\right|,
\label{3o3}
\vspace{2mm}
\end{equation}
where 
\begin{equation}
\left|\frac{V_{ub}V^{*}_{cb}}{V_{ud}V^{*}_{cd}}\right|=
\frac{A^2 \lambda^4}{1-\lambda^2/2}\sqrt{\frac{\rho^2
+\eta^2}{(1+A^2 \lambda^4 \rho)^2+A^4 \lambda^8\eta^2}}.
\label{3p}
\vspace{2mm}
\end{equation}

Next we consider $\Lambda_c\rightarrow p \rho^0 (\omega)$. Defining
\begin{equation}
\langle \rho^0 p |O^d_1 |\Lambda_c \rangle = \langle 
\rho^0 |(\bar{d}d)|0\rangle \langle p | (\bar{u} c) |\Lambda_c \rangle 
\equiv  T_2,
\label{3q1}
\vspace{2mm}
\end{equation}
we have
\begin{equation}
t^{p}_{\rho}=(c'_1+\frac{1}{N_c} c'_2)T_2.
\label{3q2}
\vspace{2mm}
\end{equation}

After evaluating $t_{\omega}^p$ and
the penguin diagram contributions we obtain the
following results,
\begin{equation}
(\alpha e^{i\delta_\alpha})^p=-1,
\label{3r1}
\vspace{2mm}
\end{equation}
\begin{equation}
(\beta e^{i\delta_\beta})^p=
\frac{c'_4+\frac{1}{N_c}c'_3}
{(2+\frac{1}{N_c})c'_3+(1+\frac{2}{N_c})c'_4+2(c'_5+\frac{1}{N_c}c'_6)},
\label{3r2}
\vspace{2mm}
\end{equation}
\begin{equation}
(r' e^{i\delta_q})^p=\frac{(2+\frac{1}{N_c})c'_3+
(1+\frac{2}{N_c})c'_4+2(c'_5+\frac{1}{N_c}c'_6)}{c'_1+\frac{1}{N_c}c'_2}
\left|\frac{V_{ub}V^{*}_{cb}}{V_{ud}V^{*}_{cd}}\right|.
\label{3r3}
\vspace{2mm}
\end{equation}

For the decay channel $D^0 \ra \phi  \rho^0 (\omega) \ra\phi \pi^+\pi^-$
the operators $O_1^s$ and $O_2^s$ contribute to the decay matrix elements.
If we define
\begin{equation}
\langle \rho^0 \phi |O^s_1 | D^0 \rangle = \langle 
\phi |(\bar{s}s)|0 \rangle \langle \rho^0 | (\bar{u} c) |D^0 \rangle 
\equiv  T_3,
\label{3s1}
\vspace{2mm}
\end{equation}
we have
\begin{equation}
t^{\phi}_{\rho}=(c'_1+\frac{1}{N_c} c'_2)T_3,
\label{3s2}
\vspace{2mm}
\end{equation}
and
\begin{equation}
(\alpha e^{i\delta_\alpha})^\phi=1,
\label{3t1}
\vspace{2mm}
\end{equation}
\begin{equation}
(\beta e^{i\delta_\beta})^\phi=1,
\label{3t2}
\vspace{2mm}
\end{equation}
\begin{equation}
(r' e^{i\delta_q})^\phi=-\frac{c'_3+\frac{1}{N_c}c'_4+c'_5+\frac{1}{N_c}c'_6}
{c'_1+\frac{1}{N_c}c'_2}
\left|\frac{V_{ub}V^{*}_{cb}}{V_{us}V^{*}_{cs}}\right|,
\label{3t3}
\vspace{2mm}
\end{equation}
where
\begin{equation}
\left|\frac{V_{ub}V^{*}_{cb}}{V_{us}V^{*}_{cs}}\right|=
\frac{A^2 \lambda^4 \sqrt{\rho^2+\eta^2}}{1-\lambda^2/2}.
\label{3u}
\vspace{2mm}
\end{equation}

For the decay channels $D^0 \ra \eta  \rho^0 (\omega) \ra\eta \pi^+\pi^-$
and $D^0 \ra \eta'  \rho^0 (\omega) \ra\eta' \pi^+\pi^-$, things become
a little complicated. It is known that $\eta$
and $\eta'$ have both $\bar u u +\bar d d$ and $\bar s s$ components. The
decay constants, $f_{\eta (\eta')}^u$ and  $f_{\eta (\eta')}^s$, defined as
\begin{equation}
\langle 0 |\bar u \gamma_\mu \gamma_5 u | \eta (\eta') \rangle = 
i f_{\eta (\eta')}^u p_\mu, \;\;\;\;\;
\langle 0 |\bar s \gamma_\mu \gamma_5 s | \eta (\eta') \rangle = 
i f_{\eta (\eta')}^s p_\mu,
\label{3v}
\vspace{2mm}
\end{equation}
are different. After straightforward derivations we have
\begin{equation}
(\alpha e^{i\delta_\alpha})^{\eta (\eta')}=1,
\label{3w1}
\vspace{2mm}
\end{equation}
\begin{equation}
(\beta e^{i\delta_\beta})^{\eta (\eta')}=1,
\label{3w2}
\vspace{2mm}
\end{equation}
\begin{equation}
(r' e^{i\delta_q})^{\eta (\eta')}=-\frac{2 f_{\eta (\eta')}^u +
f_{\eta (\eta')}^s}{f_{\eta (\eta')}^u - f_{\eta (\eta')}^s}
\frac{c'_3+\frac{1}{N_c}c'_4-c'_5-\frac{1}{N_c}c'_6}
{c'_1+\frac{1}{N_c}c'_2}
\left|\frac{V_{ub}V^{*}_{cb}}{V_{us}V^{*}_{cs}}\right|.
\label{3w3}
\vspace{2mm}
\end{equation}
In the derivations of Eqs.(\ref{3w1}, \ref{3w2},\ref{3w3}) 
we have made the approximation that $V_{ub}V^{*}_{cb}/V_{ud}V^{*}_{cd}=
-V_{ub}V^{*}_{cb}/V_{us}V^{*}_{cs}$. It is noted that the minus signs 
associated with $c'_5$ and $c'_6$ in Eq.(\ref{3w3})
arise because $\eta (\eta')$ are pseudoscalar
mesons. Since the imaginary part of $c'_3 (c'_4)$ is the same as
that of $c'_5 (c'_6)$, $\delta_q$ is zero. This leads to the strong phase,
$\delta$, being zero, in combination with Eqs.(\ref{3w1}, \ref{3w2}).

The decay constants $f_{\eta (\eta')}^u$ and  $f_{\eta (\eta')}^s$ were
calculated phenomenologically in Ref.\cite{stech}, based on the assumption
that the decay constants in the quark flavor basis follow the pattern of 
particle state mixing. It was found that 
\begin{equation}
f_{\eta}^u=78{\rm MeV},\;\;f_{\eta}^s=-112{\rm MeV},\;\;
f_{\eta'}^u=63{\rm MeV},\;\;f_{\eta'}^s=137{\rm MeV}.
\label{3x}
\vspace{2mm}
\end{equation}

For the decay process $D^+ \ra \pi^+ \rho^0 (\omega) \ra \pi^+ \pi^+\pi^-$,
two kinds of matrix element products are involved after factorization, i.e., 
$\langle \rho^0 (\omega) |(\bar{d}d)|0 \rangle \langle \pi^+ | (\bar{u} c) 
|D^+ \rangle$ and  
$\langle \pi^+ |(\bar{u}d)|0 \rangle \langle \rho^0 (\omega) | (\bar{d} c) 
|D^+ \rangle$. These two quantities cannot be related to each other by
symmetry. Therefore, we have to evaluate them in some phenomenological
quark models and hence more uncertainties are involved. Similarly for
$D^0 \ra \pi^0 \rho^0 (\omega) \ra \pi^0 \pi^+\pi^-$ we have to evaluate
$\langle \rho^0 (\omega) |(\bar{d}d)|0 \rangle \langle \pi^0 | (\bar{u} c) 
|D^0 \rangle$ and  
$\langle \pi^0 |(\bar{d}d)|0 \rangle \langle \rho^0 (\omega) | (\bar{u} c) 
|D^0 \rangle$ seperately.

The matrix elements for $D\ra X$ and $D\ra X^*$ ($X$ and $X^*$ 
denote pseudoscalar and vector mesons, respectively) can be decomposed 
as\cite{stech2},
\begin{eqnarray}
\langle X|J_\mu |D \rangle &=& \left(p_D +p_X -\frac{m_D^2-m_X^2}{k^2}
k \right)_\mu
F_1 (k^2)+\frac{m_D^2-m_X^2}{k^2}k_\mu F_0 (k^2),
\label{3y1}\\
\langle X^*|J_\mu |D \rangle &=& \frac{2}{m_D+m_{X^*}}\epsilon_{\mu\nu\rho
\sigma}\epsilon^{*\nu}p_{D}^{\rho}p_{X^*}^{\sigma} V(k^2)+i\left[
\epsilon^*_\mu (m_D+m_{X^*})A_1(k^2)\right.\nn\\
&&\left.-\frac{\epsilon\cdot k}{m_D+m_{X^*}}(p_D+p_{X^*})_\mu A_2(k^2)
-\frac{\epsilon\cdot k}{k^2}2m_{X^*}k_\mu A_3(k^2)\right]\nn\\
&&+i\frac{\epsilon\cdot k}{k^2}2m_{X^*}k_\mu  A_0(k^2),
\label{3y2}
\vspace{2mm}
\end{eqnarray}
where $J_\mu$ is the weak current, $k=p_D-p_{X(X^*)}$ and $\epsilon_\mu$ is the
polarization vector of $X^*$. The form factors 
satisfy the relations $F_1(0)=F_0(0)$, $A_3(0)=A_0(0)$ and $A_3(k^2)=
\frac{m_D+m_{X^*}}{2m_{X^*}}A_1(k^2)-\frac{m_D-m_{X^*}}{2m_{X^*}}A_2(k^2)$.

Using the decomposition in Eqs.(\ref{3y1})(\ref{3y2}), we have for 
$D^+ \ra \pi^+ \rho^0 (\omega)$,
\begin{equation}
t_{\rho}^{\pi^+}=-\sqrt 2 m_D |\vec{p}_\rho|
\left[\left(c'_1+\frac{1}{N_c}c'_2\right)f_\rho F_1(m_\rho^2)
+\left(c'_2+\frac{1}{N_c}c'_1\right)f_\pi A_0(m_\pi^2)\right],
\label{3z}
\vspace{2mm}
\end{equation}
where $f_\rho$ and $f_\pi$ are the decay constants of the $\rho$ and $\pi$, 
respectively, and $\vec{p}_\rho$ is the three momentum of the $\rho$.

It can be shown that $t_{\omega}^{\pi^+}=-t_{\rho}^{\pi^+}$. After calculating
the penguin operator contributions, we have
\begin{equation}
(\alpha e^{i\delta_\alpha})^{\pi^+}=-1,
\label{3aa1}
\vspace{2mm}
\end{equation}
\begin{equation}
(\beta e^{i\delta_\beta})^{\pi^+}=\frac{[f_\rho F_1(m_\rho^2)
-f_\pi A_0(m_\pi^2)](c'_4+\frac{1}{N_c} c'_3)-\frac{2m_\pi^2 
f_\pi A_0(m_\pi^2)}{(m_c+m_d)(m_u+m_d)}(c'_6+\frac{1}{N_c} c'_5)}{x},
\label{3aa2}
\vspace{2mm}
\end{equation}
\begin{equation}
(r' e^{i\delta_q})^{\pi^+}=\frac{x}{[f_\rho F_1(m_\rho^2)+\frac{1}{N_c}
f_\pi A_0(m_\pi^2)]c'_1+[\frac{1}{N_c}f_\rho F_1(m_\rho^2)+f_\pi A_0(m_\pi^2)]
c'_2}\left|\frac{V_{ub}V^{*}_{cb}}{V_{ud}V^{*}_{cd}}\right|,
\label{3aa3}
\vspace{2mm}
\end{equation}
where $x$ is defined as
\begin{eqnarray}
x&=&\left[2f_\rho F_1(m_\rho^2)+\frac{f_\rho F_1(m_\rho^2)+f_\pi A_0(m_\pi^2)}
{N_c}\right]c'_3+\left[\frac{2f_\rho F_1(m_\rho^2)}{N_c}+f_\rho F_1(m_\rho^2)
\right.\nn\\
&&\left.+f_\pi A_0(m_\pi^2)\right]c'_4
+2\left[f_\rho F_1(m_\rho^2)-\frac{m_\pi^2 f_\pi A_0(m_\pi^2)}{N_c
(m_c+m_d)(m_u+m_d)}\right]c'_5\nn\\
&&+2\left[\frac{f_\rho F_1(m_\rho^2)}{N_c}
-\frac{m_\pi^2 f_\pi A_0(m_\pi^2)}{(m_c+m_d)(m_u+m_d)}\right]c'_6.
\label{3aa4}
\vspace{2mm}
\end{eqnarray}

We can consider the process 
$D^0 \ra \pi^0 \rho^0 (\omega) \ra \pi^0 \pi^+\pi^-$ in the same way. We 
find
\begin{equation}
(\alpha e^{i\delta_\alpha})^{\pi^0}=-\frac{f_\rho F_1(m_\rho^2)
-f_\pi A_0(m_\pi^2)}{f_\rho F_1(m_\rho^2)+f_\pi A_0(m_\pi^2)},
\label{3bb1}
\vspace{2mm}
\end{equation}
\begin{equation}
(\beta e^{i\delta_\beta})^{\pi^0}=\frac{[f_\rho F_1(m_\rho^2)
+f_\pi A_0(m_\pi^2)](c'_4+\frac{1}{N_c} c'_3)-\frac{2m_\pi^2 
f_\pi A_0(m_\pi^2)}{(m_c+m_d)(m_u+m_d)}(c'_6+\frac{1}{N_c} c'_5)}{x},
\label{3bb2}
\vspace{2mm}
\end{equation}
\begin{equation}
(r' e^{i\delta_q})^{\pi^0}=\frac{x}{[f_\rho F_1(m_\rho^2)+f_\pi A_0(m_\pi^2)]
(c'_1+\frac{1}{N_c}c'_2)}\left|\frac{V_{ub}V^{*}_{cb}}{V_{ud}V^{*}_{cd}}\right|.
\label{3bb3}
\vspace{2mm}
\end{equation}

In Eqs.(\ref{3aa1})-(\ref{3bb3}) the form factors $F_1 (m_\rho^2)$
and $A_0(m_\pi^2)$ depend on the inner structure of the hadrons. Under the 
nearest pole dominance assumption, the $k^2$ dependence of these form factors
are
\begin{equation}
F_1(k^2)=\frac{h_1}{1-k^2/m_1^2},\;\;\; A_0(k^2)=\frac{h_{A_0}}
{1-k^2/m_{A_0}^2},
\label{3cc}
\vspace{2mm}
\end{equation}
where $m_1=2.01$GeV and $m_{A_0}=1.87$GeV\cite{stech2} and $h_1$ and $h_{A_0}$
are given by the overlap integrals of the hadronic wave functions of $D$
and $X(X^*)$\cite{stech2, gh}.

Having obtained the expressions for $\alpha e^{i\delta_\alpha}$,
$\beta e^{i\delta_\beta}$ and $r' e^{i\delta_q}$, 
for different decay processes, we may
substitute them into Eq.(\ref{3j}) to obtain 
$(r {\rm sin}\delta)$ and $(r {\rm cos}\delta)$ for each channel.
Then, in combination with with Eqs.(\ref{3l1}) and (\ref{3l2}), 
the CP-violating asymmetries $a$ can be obtained from Eq.(\ref{3d}).

\vspace{0.2in}
{\bf III.3 Numerical results} 
\vspace{0.2in}

In the numerical calculations, we have several parameters: $q^2$, $N_c$,
and the CKM matrix elements in the Wolfenstein parametrization.
As mentioned in Section II, the value
of $q^2$ is conventionally 
chosen to be in the range $0.3 < q^2/m_{c}^2 < 0.5$. 
For the CKM matrix elements, which should be determined from experiment, 
we use $\lambda=0.221$, $\eta=0.34$ and $\rho=-0.12$ as in Ref.\cite{ga}.

The value of the effective $N_c$ should also be determined by experiment.
Since the hadronization information is included in $N_c$, the value of $N_c$
may be different for different decay channels. Furthermore, since the 
color-octet contribution associated with each operator in the Hamiltonian 
(\ref{2a}) can vary, the effective $N_c$ in the Fierz transformation
for each operator may be different. In general, nonfactorizable effects can
be absorbed into the effective parameters $a_i^{\rm eff}$ after the
Fierz transformation,
\begin{equation}
a_{2i}^{\rm eff}=c'_{2i}+\frac{1}{(N_c)_{2i}}c'_{2i-1},\;\;\;
a_{2i-1}^{\rm eff}=c'_{2i-1}+\frac{1}{(N_c)_{2i-1}}c'_{2i},\;\;
(i=1,2,3),
\label{3dd}
\vspace{2mm}
\end{equation}
where
\begin{equation}
\frac{1}{(N_c)_i}\equiv \frac{1}{3}+\xi_i \;\; (i=1,...,6),
\label{3ee}
\vspace{2mm}
\end{equation}
with $\xi_i$ being the nonfactorizable effects, which may be different for
each operator. However, since we do not have enough information about
the operator dependence of $\xi_i$, we assume $\xi_i$ is universal for
each operator\cite{ali} and hence for each operator we use the same effective
$N_c$ ($=(N_c)_i$). 

In the numerical calculations, it is found that
for a fixed $N_c$ there is a maximum point, $a_{\rm max}$, 
for the CP violating parameter $a$, when the invariant mass of the 
$\pi^+\pi^-$ pair is in the vicinity of the $\omega$ resonance.
We have calculated $a_{\rm max}$ in the range $N_c > 0$ for different
decay channels. In the calculations we use the following
two sets of form factors\cite{stech2}:
$${\rm Set \;1}: h_1=0.69,\;\; h_{A_0}=0.67,\;\;h_{V}=1.23,\;\;
h_{A_1}=0.78,\;\;h_{A_2}=0.92,$$
$${\rm Set \;2}: h_1=0.78,\;\; h_{A_0}=0.77,\;\;h_{V}=1.55,\;\;
h_{A_1}=0.98,\;\;h_{A_2}=1.27.$$
The above two sets of parameters correspond to taking the average 
transverse momentum of the constituents in the meson to be 400MeV or 500MeV, 
respectively\cite{stech2}. The $k^2$ dependence of $A_1 (k^2)$,
$A_2 (k^2)$ and $V(k^2)$ are the same as in Eq.(\ref{3cc}).

The numerical results show that for 
$D^+ \ra \rho^+ \rho^0 (\omega) \ra \rho^+ \pi^+\pi^-$, in the whole range
$N_c > 0$, we have $a_{\rm max} \leq 3\times 10^{-4}$, which is small.
For $D^0 \ra \eta  \rho^0 (\omega) \ra\eta \pi^+\pi^-$ and  
$D^0 \ra \eta'  \rho^0 (\omega) \ra\eta' \pi^+\pi^-$, from Eqs.(\ref{3w1},
\ref{3w2}, \ref{3w3}) 
it can be seen that the strong phase $\delta$ is zero. Therefore, we
do not have CP violation in these decays in our approach. However, for other
processes there is a small range of $N_c$ in which we may have large
$a_{\rm max}$ ($\geq 1\%$).

For $D^0 \ra \phi \rho^0 (\omega) \ra \phi \pi^+\pi^-$, the range of $N_c$
for $a_{\rm max}\geq 1\%$ is $1.98\leq N_c \leq 1.99$, while for 
$\Lambda_c \ra p \rho^0 (\omega) \ra p \pi^+\pi^-$ the range is
$1.95\leq N_c \leq 2.02$. For 
$D^+ \ra \pi^+ \rho^0 (\omega) \ra \pi^+ \pi^+\pi^-$ we find that for the
first set of form factors when $N_c \geq 56$,
$a_{\rm max} \geq 1\%$, while for the second set of form factors 
when $N_c \geq 136$, $a_{\rm max} \geq 1\%$, 
in the range $0.3 \leq q^2/m_c^2 \leq 0.5$. For 
$D^0 \ra \pi^0 \rho^0 (\omega) \ra \pi^0 \pi^+\pi^-$ we find that when
$1.98 \leq N_c \leq 1.99$ we have $a_{\rm max} \geq 1\%$ in the range
$0.3 \leq q^2/m_c^2 \leq 0.5$ for both sets of form factors.
  
The above ranges for $N_c$ were obtained by the requirement that  
we have large CP violation in this range. However, whether $N_c$
can be in this range should be determined by the experimental data for
the branching ratio of each decay channel. Usually the 
decay rate for $D\ra f \rho^0$
is determined primarily by the tree operators, $O_1$ and $O_2$,
which  are related to $t_\rho^f$. In fact, the reason why we can find 
large CP violation in some range of $N_c$, is that in this range
$t_\rho^f$ becomes small enough so that $r'$, and hence $r$, becomes large
(see Eqs.(\ref{3f}, \ref{3g})). However, if  $t_\rho^f$ is too small the
decay rate it yields 
for $D\ra f \rho^0$ may be smaller than the experimental
data. In such a case, the range of $N_c$ in which we could have large CP
violation will be excluded by the data. 

The decay widths for nonleptonic decays of $D$-meson can be calculated
straightforwardly in the quark model of Refs.\cite{stech2, gh}. 
Since we are considering
the range for $N_c$ in which $t_\rho^f$ is small, we have to take into account
the penguin contributions $p_\rho^f$ as well when we calculate the decay
widths. In the calculations of the decay width for $D^0 \ra \phi \rho^0$ 
we use $f_\phi=
237$MeV. We find that for the first set of form factors the
branching ratio is smaller than $2.3\times 10^{-8}$ and for the second set 
the branching ratio is smaller than $3.6\times 10^{-8}$ in the range 
$1.98 \leq N_c \leq 1.99$. The dependence of the branching ratio on $q^2/m_c^2$
is negligible. These branching ratios are much 
smaller than  the experimental data $(6\pm 3)\times 10^{-4}$\cite{data} which
corresponds to $1.31 \leq N_c \leq 1.53$ ($1.41 \leq N_c \leq 1.60$) 
for the first (second) set of form factors. Similarly, when
$N_c \geq 56$ the branching ratio for $D^+ \ra \pi^+ \rho^0$ is smaller than
$1.0\times 10^{-5}$, while the experimental data is $(1.05\pm 0.31)\times 
10^{-3}$\cite{data} corresponding to $2.1 \leq N_c \leq 2.9$ 
($2.5 \leq N_c \leq 3.4$) 
for the first
(second) set of form factors. Therefore, we cannot have large CP violation in
$D^0 \ra \phi \rho^0 (\omega) \ra \phi \pi^+\pi^-$ and 
$D^+ \ra \pi^+ \rho^0 (\omega) \ra \pi^+ \pi^+\pi^-$.
 
However, for the decay processes 
$D^0 \ra \pi^0 \rho^0$ and $\Lambda_c \ra p \rho^0$ there are no 
experimental data at present\cite{data}. 
Therefore, there is still a possibility that
$N_c$ could be in the range required for large CP violation for 
$D^0 \ra \pi^0 \rho^0 (\omega) \ra \pi^0 \pi^+\pi^-$ or
$\Lambda_c \ra p \rho^0 (\omega) \ra p \pi^+\pi^-$. 
The decay width for $D^0 \ra \pi^0 \rho^0$ is calculated in the same way
and we find that for the first set of of form factors the
branching ratio is $1.4 (1.7)\times 10^{-8}$, while for the second set 
the branching ratio is $1.8 (2.1)\times 10^{-8}$ for 
$N_c =1.98 (1.99)$. This prediction is almost independent of 
$q^2/m_c^2$.

The branching ratio for  $\Lambda_c \ra p \rho^0$ can be calculated
with the same method as that 
in Ref.\cite{ga}, where we worked in the heavy quark limit
$m_c \ra \infty$ and used the diquark model hadronic wavefunctions for both
the heavy baryon, $\Lambda_c$, and the proton, $p$. 
As in the neutron case, in the 
diquark model the Clebsch-Gordan coefficient of the $u[ud]$ component 
($[ud]$ is the scalar
diquark) is also $1/\sqrt{2}$ for the proton\cite{kroll}. We find that for
$N_c =1.95 (2.02)$ the
branching ratio for  $\Lambda_c \ra p \rho^0$ is $7.9 (7.5) 
\times 10^{-9}$ for  $b=1.77GeV^{-1}$, corresponding to 
$\langle k_{\perp}^{2}\rangle$$^{\frac{1}{2}}$ = 400 MeV, and
$6.9 (7.0) \times 10^{-9}$ for $b=1.18GeV^{-1}$, corresponding to 
$\langle k_{\perp}^{2}\rangle$$^{\frac{1}{2}}$ = 600 MeV, where 
$\langle k_{\perp}^{2}\rangle$ is the average transverse momentum of the 
$c$ quark in the $\Lambda_c$. 
Again the branching ratio is very insensitive to $q^2/m_c^2$.

In Tables 1-5 we list numerical results for $a_{\rm max}$ and $Br(H_c \ra
f \rho^0)$ for various processes, with different values of $N_c$
and $q^2/m_c^2$. It should be 
noted that $Br(H_c \ra f \rho^0)$ is almost same for
$q^2/m_c^2=0.3$ and $q^2/m_c^2=0.5$. Table 1 shows explicitly that,
for $D^+ \ra \rho^+ \rho^0 (\omega) \ra \rho^+ \pi^+\pi^-$, $a_{\rm max}$
is at most $\sim 10^{-4}$ no matter what $N_c$ is. From Tables 2 and 3
we can see that in the region of $N_c$ allowed by the experimental data,
$a_{\rm max}$ is of the order $10^{-4}$ for 
$D^0 \ra \phi  \rho^0 (\omega) \ra\phi \pi^+\pi^-$ and
$D^+ \ra \pi^+ \rho^0 (\omega) \ra \pi^+ \pi^+\pi^-$. It can also be seen
explicitly from Tables 4 and 5 that there is a range for $N_c$ in which
$a_{\rm max}$ may be bigger than 1\% for  
$D^0 \ra \pi^0 \rho^0 (\omega) \ra \pi^0 \pi^+\pi^-$ and
$\Lambda_c \ra p \rho^0 (\omega) \ra p \pi^+\pi^-$. 

\begin{table}
\caption{Values of $Br(D^+ \ra \rho^+ \rho^0)$ with the first (second) set
of form factors and $a_{\rm max}$ for 
$D^+ \ra \rho^+ \rho^0 (\omega) \ra \rho^+ \pi^+\pi^-$, with 
$q^2/m_c^2=0.3(0.5)$}
\begin{center}
\begin{tabular}{lccc}
\hline
\hline
$N_c$&$Br(D^+ \ra \rho^+ \rho^0)$  &$a_{\rm max}$   \\ 
\hline
0.5&6.3(9.6)$\times 10^{-2}$&2.0(1.8)$\times 10^{-4}$\\
\hline
1.0&2.5(3.8)$\times 10^{-2}$&1.4(1.3)$\times 10^{-4}$\\
\hline
1.5&1.6(2.5)$\times 10^{-2}$&1.1(0.96)$\times 10^{-4}$\\
\hline
2.0&1.3(1.9)$\times 10^{-2}$&8.7(7.7)$\times 10^{-5}$\\
\hline
3.0&0.95(1.5)$\times 10^{-2}$&6.1(5.3)$\times 10^{-5}$\\
\hline
\hline
\end{tabular}
\end{center}
\end{table}

\begin{table}
\caption{Values of $Br(D^0 \ra \phi \rho^0)$ and $a_{\rm max}$ for 
$D^0 \ra \phi \rho^0 (\omega) \ra \phi \pi^+\pi^-$, with $q^2/m_c^2=0.3(0.5)$}
\begin{center}
\begin{tabular}{lccc}
\hline
\hline
&First set of form factors  &\\
\hline
\hline
$N_c$&$Br(D^0 \ra \phi \rho^0)$  &$a_{\rm max}$   \\ 
\hline
1.31&9.0$\times 10^{-4}$&1.2(1.3)$\times 10^{-4}$\\
\hline
1.36&7.2$\times 10^{-4}$&1.3(1.3)$\times 10^{-4}$\\
\hline
1.41&5.7$\times 10^{-4}$&1.4(1.4)$\times 10^{-4}$\\
\hline
1.46&4.4$\times 10^{-4}$&1.5(1.5)$\times 10^{-4}$\\
\hline
1.53&3.0$\times 10^{-4}$&1.6(1.6)$\times 10^{-4}$\\
\hline
\hline
&Second set of form factors  &\\
\hline
\hline
$N_c$&$Br(D^0 \ra \phi \rho^0)$  &$a_{\rm max}$   \\ 
\hline
1.41&8.9$\times 10^{-4}$&1.4(1.4)$\times 10^{-4}$\\
\hline
1.46&6.9$\times 10^{-4}$&1.5(1.5)$\times 10^{-4}$\\
\hline
1.51&5.3$\times 10^{-4}$&1.6(1.6)$\times 10^{-4}$\\
\hline
1.56&4.0$\times 10^{-4}$&1.7(1.7)$\times 10^{-4}$\\
\hline
1.60&3.0$\times 10^{-4}$&1.8(1.8)$\times 10^{-4}$\\
\hline
\hline
\end{tabular}
\end{center}
\end{table}

\begin{table}
\caption{Values of $Br(D^+ \ra \pi^+ \rho^0)$ and $a_{\rm max}$ for 
$D^+ \ra \pi^+ \rho^0 (\omega) \ra \pi^+ \pi^+\pi^-$, 
with $q^2/m_c^2=0.3(0.5)$}
\begin{center}
\begin{tabular}{lccc}
\hline
\hline
&First set of form factors  &\\
\hline
\hline
$N_c$&$Br(D^+ \ra \pi^+ \rho^0)$  &$a_{\rm max}$   \\ 
\hline
2.1&1.4$\times 10^{-3}$&-3.0(-3.0)$\times 10^{-4}$\\
\hline
2.5&9.8$\times 10^{-4}$&-3.9(-4.0)$\times 10^{-4}$\\
\hline
2.9&7.3$\times 10^{-4}$&-4.9(-4.9)$\times 10^{-4}$\\
\hline
\hline
&Second set of form factors  &\\
\hline
\hline
$N_c$&$Br(D^+ \ra \pi^+ \rho^0)$  &$a_{\rm max}$   \\ 
\hline
2.5&1.3$\times 10^{-3}$&-4.0(-4.0)$\times 10^{-4}$\\
\hline
3.0&9.2$\times 10^{-4}$&-5.2(-5.1)$\times 10^{-4}$\\
\hline
3.4&7.3$\times 10^{-4}$&-6.1(-6.0)$\times 10^{-4}$\\
\hline
\hline
\end{tabular}
\end{center}
\end{table}

\begin{table}
\caption{Values of $Br(D^0 \ra \pi^0 \rho^0)$ with the first (second) set
of form factors and $a_{\rm max}$ for 
$D^0 \ra \pi^0 \rho^0 (\omega) \ra \pi^0 \pi^+\pi^-$, 
with $q^2/m_c^2=0.3(0.5)$}
\begin{center}
\begin{tabular}{lccc}
\hline
\hline
$N_c$&$Br(D^0 \ra \pi^0 \rho^0)$ &$a_{\rm max}$ (Set 1)
&$a_{\rm max}$ (Set 2)   \\ 
\hline
0.5&2.1(2.8)$\times 10^{-2}$&1.0(0.94)$\times 10^{-4}$
&1.1(0.93)$\times 10^{-4}$\\
\hline
1.0&2.3(3.0)$\times 10^{-3}$&9.5(8.2)$\times 10^{-5}$
&9.3(8.1)$\times 10^{-5}$\\
\hline
1.5&2.5(3.3)$\times 10^{-4}$&9.7(8.7)$\times 10^{-5}$
&9.3(8.4)$\times 10^{-5}$\\
\hline
1.9&4.8(6.2)$\times 10^{-5}$&-7.1(-8.3)$\times 10^{-4}$
&-7.4(-8.6)$\times 10^{-4}$\\
\hline
1.98&1.4(1.8)$\times 10^{-8}$&-1.4(-1.6)$\times 10^{-2}$
&-1.5(-1.7)$\times 10^{-2}$\\
\hline
1.99&1.7(2.1)$\times 10^{-8}$&1.4(1.6)$\times 10^{-2}$
&1.4(1.7)$\times 10^{-2}$\\
\hline
2.1&7.3(9.4)$\times 10^{-6}$&7.2(8.1)$\times 10^{-4}$
&7.5(8.3)$\times 10^{-4}$\\
\hline
2.5&1.0(1.3)$\times 10^{-4}$&2.5(2.8)$\times 10^{-4}$
&2.7(2.9)$\times 10^{-4}$\\
\hline
3.0&2.8(3.6)$\times 10^{-4}$&2.0(2.1)$\times 10^{-4}$
&2.1(2.1)$\times 10^{-4}$\\
\hline
10.0&1.6(2.0)$\times 10^{-3}$&1.5(1.4)$\times 10^{-4}$
&1.5(1.4)$\times 10^{-4}$\\
\hline
\hline
\end{tabular}
\end{center}
\end{table}

\begin{table}
\caption{Values of $Br(\Lambda_c \ra p \rho^0)$ with 
$\langle k_{\perp}^{2}\rangle$$^{\frac{1}{2}}$ = 400 MeV (600MeV)
and $a_{\rm max}$ for 
$\Lambda_c \ra p \rho^0 (\omega) \ra p \pi^+\pi^-$, with 
$q^2/m_c^2=0.3(0.5)$}
\begin{center}
\begin{tabular}{lccc}
\hline
\hline
$N_c$&$Br(\Lambda_c \ra p \rho^0)$  &$a_{\rm max}$   \\ 
\hline
0.5&2.2(1.9)$\times 10^{-4}$&2.0(1.8)$\times 10^{-4}$\\
\hline
1.0&2.4(2.1)$\times 10^{-5}$&3.3(3.0)$\times 10^{-4}$\\
\hline
1.5&2.6(2.3)$\times 10^{-6}$&7.3(6.7)$\times 10^{-4}$\\
\hline
1.9&4.9(4.3)$\times 10^{-8}$&3.9(4.1)$\times 10^{-3}$\\
\hline
1.95&7.9(6.9)$\times 10^{-9}$&1.0(1.1)$\times 10^{-2}$\\
\hline
2.02&7.5(7.0)$\times 10^{-9}$&-1.1(-1.0)$\times 10^{-2}$\\
\hline
2.1&7.5(6.5)$\times 10^{-8}$&-3.4(-3.1)$\times 10^{-3}$\\
\hline
2.5&1.1(0.92)$\times 10^{-6}$&-8.0(-7.4)$\times 10^{-4}$\\
\hline
3.0&2.8(2.5)$\times 10^{-6}$&-4.3(-4.0)$\times 10^{-4}$\\
\hline
10.0&1.6(1.4)$\times 10^{-5}$&-1.1(-1.1)$\times 10^{-4}$\\
\hline
\hline
\end{tabular}
\end{center}
\end{table}

In Fig.1 we plot the numerical values of the CP-violating 
asymmetries, $a$, for $D^0 \ra \pi^0 \rho^0 (\omega) \ra \pi^0 \pi^+\pi^-$
with $N_c=1.99$ and $q^2/m_c^2=0.3$, 0.5 (for $N_c=1.98$ we have similar 
results) as a function of the invariant mass of the $\pi^+\pi^-$
pair. It should be noted that in Fig.1 we used the first set of
form factors. The results for the second set change very little. 
In  Fig.2 we plot the results for 
$\Lambda_c \ra p \rho^0 (\omega) \ra p \pi^+\pi^-$, with $N_c=2.02$
and $q^2/m_c^2=0.3$, 0.5 (for $N_c=1.95$ we have similar 
results). In both of these plots we find that we can 
have $a_{\rm max} \geq 1\%$. \\

\begin{figure}[htb]
\begin{center}
\leavevmode
\epsfxsize = 470pt
\epsfysize = 370pt
\epsfbox[1 240 578 702]{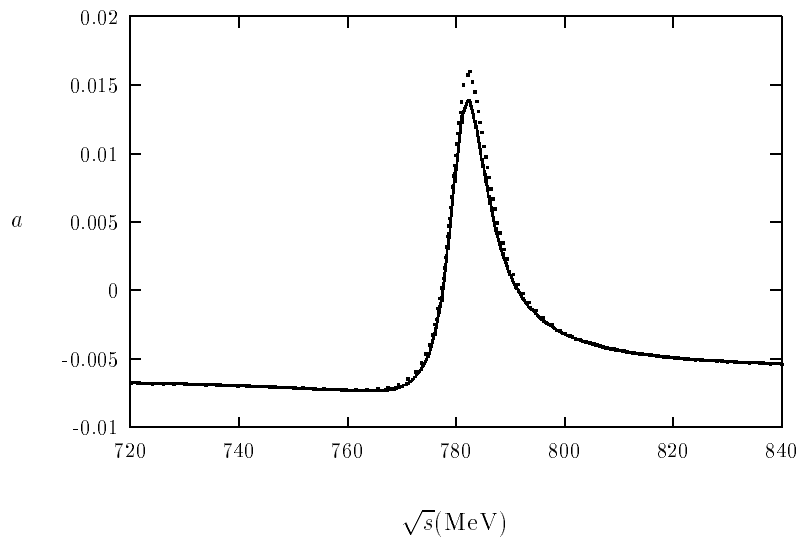}
\end{center}
\vspace{-8cm}
\caption{The CP-violating asymmetry for  
$D^0 \rightarrow \pi^0 \rho^0 (\omega) \rightarrow \pi^0 \pi^+\pi^-$
(with $N_c=1.99$) as a function of the invariant mass of the $\pi^+\pi^-$
pair. The solid (dotted) line is for $q^2/m_{c}^{2}=0.3$ (0.5).}
\label{FIG1}  
\end{figure}

\vspace{0.2cm}

\begin{figure}[htb]
\begin{center}
\leavevmode
\epsfxsize = 470pt
\epsfysize = 370pt
\epsfbox[1 240 578 702]{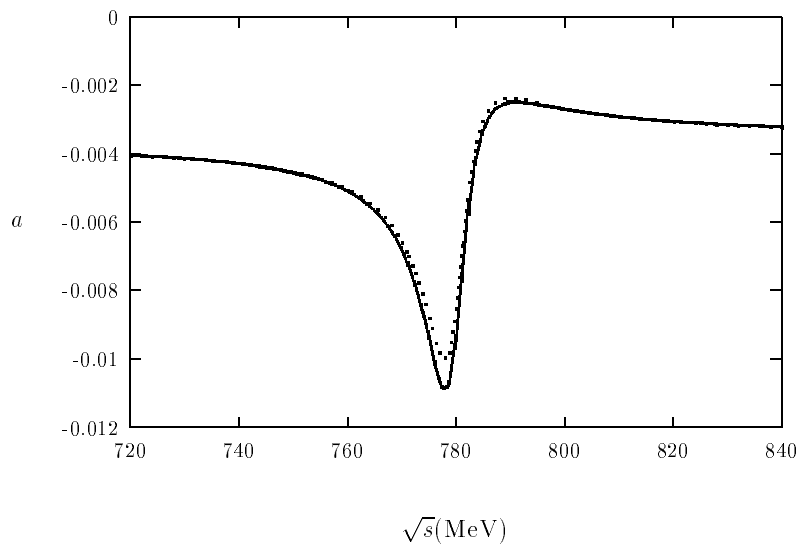}
\end{center}
\vspace{-8cm}
\caption{The CP-violating asymmetry for 
$\Lambda_c \ra p \rho^0 (\omega) \ra p \pi^+\pi^-$,
with $N_c=2.02$. The solid (dotted) line is for $q^2/m_{c}^{2}=0.3$ (0.5).}
\label{FIG2}  
\end{figure}

\vspace{0.2in}
{\large\bf IV. Summary and discussions}
\vspace{0.2in}

The aim of the present work was to look for possibilities of large CP
violation in charmed meson or baryon nonleptonic decays, 
$H_c \ra f \pi^+\pi^-$.
Since CP violation in the charm sector
is usually estimated to be very small (less than $10^{-3}$),
it would be fascinating to find cases where the CP violation is large
($> 1\%$). Following our previous work on 
CP violation in the b-quark system\cite{tony, ga}, 
we have studied direct CP violation in 
$D^+ \ra \rho^+ \rho^0 (\omega) \ra \rho^+ \pi^+\pi^-$,   
$D^+ \ra \pi^+ \rho^0 (\omega) \ra \pi^+ \pi^+\pi^-$,
$D^0 \ra \phi  \rho^0 (\omega) \ra\phi \pi^+\pi^-$,  
$D^0 \ra \eta  \rho^0 (\omega) \ra\eta \pi^+\pi^-$,  
$D^0 \ra \eta'  \rho^0 (\omega) \ra\eta' \pi^+\pi^-$,  
$D^0 \ra \pi^0 \rho^0 (\omega) \ra \pi^0 \pi^+\pi^-$, and
$\Lambda_c \ra p \rho^0 (\omega) \ra p \pi^+\pi^-$ via $\rho-\omega$ mixing.
The advantage of $\rho-\omega$ mixing is that the strong phase difference
is large (passing through $90^\circ$ at the $\omega$ resonance),
for some fixed $N_c$. As a result, the CP-violating asymmetry, $a$, has 
a maximum, $a_{\rm max}$, when the invariant
mass of the $\pi^+\pi^-$ pair is in the vicinity of the $\omega$ resonance.
It was found that  $a_{\rm max}$
depends strongly on the effective parameter $N_{c}$. For 
$D^+ \ra \rho^+ \rho^0 (\omega) \ra \rho^+ \pi^+\pi^-$,
$D^0 \ra \eta  \rho^0 (\omega) \ra\eta \pi^+\pi^-$ and 
$D^0 \ra \eta'  \rho^0 (\omega) \ra\eta' \pi^+\pi^-$, $a_{\rm max}$
is small over the whole range, $N_c \geq 0$.  However, for other processes 
we found that in order to have $a_{\rm max} \geq 1\%$, $N_c$ should be in 
a particular range in which the amplitude $t_\rho^f$ becomes small 
enough so that 
we can have large CP violation. This is because when $t_\rho^f$ is small 
$r$ can become large, leading to large CP violation. However, whether or not
$N_c$ can be in such a range is determined by the decay branching ratios
for $H_c \ra f \rho^0$. 

The experimental data exclude the possibility of large CP violation
in $D^+ \ra \pi^+ \rho^0 (\omega) \ra \pi^+ \pi^+\pi^-$
and $D^0 \ra \phi  \rho^0 (\omega) \ra\phi \pi^+\pi^-$.
However, since we do not have data for $D^0 \ra \pi^0 \rho^0$ and
$\Lambda_c \ra p \rho^0$ at present, it is still possible 
that we could have $a_{\rm max} \geq 1\%$ for 
$D^0 \ra \pi^0 \rho^0 (\omega) \ra \pi^0 \pi^+\pi^-$ and
$\Lambda_c \ra p \rho^0 (\omega) \ra p \pi^+\pi^-$ via $\rho-\omega$ mixing
in some small range of $N_c$. We estimated that in order to have large
CP violation, the branching ratios for $D^0 \ra \pi^0 \rho^0$ and
$\Lambda_c \ra p \rho^0$ should be around $10^{-9}\sim10^{-8}$ ($N_c=1.98$
or 1.99 for $D^0 \ra \pi^0 \rho^0$ and $N_c=1.95$
or 2.02 for $\Lambda_c \ra p \rho^0$). It will be very interesting to 
look for such large CP-violating asymmetries in the experiments
in order to get a deeper understanding of the mechanism for CP violation.
On the other hand, the smaller branching ratios 
will make the measurements more difficult. Furthermore, the study of
CP violation in $\Lambda_c$ decays may provide insight into the
baryon asymmetry phenomena required for baryogenesis. 

Our analysis can be extended straightforwardly
to say, $\Xi'_c \ra \Lambda \rho^0 (\omega) \ra \Lambda
\pi^+ \pi^-$, and also $D_s \ra K^* \rho^0 (\omega) \ra
K^* \pi^+ \pi^-$, if we assume SU(3) flavor symmetry.

In the calculations of CP violating asymmetry parameters we need the 
Wilson coefficients for the tree and penguin operators at the
decay scale $m_c$. We calculated the six Wilson coefficients 
to the next-to-leading order by applying
the formalism developed in \cite{buras2, ciuchini} and the relevant anomalous
dimension matrix elements. Since we only considered strong penguin operators,
and since the strong interaction is independent of flavor, 
the relevant formulas
in \cite{buras2, ciuchini} can be applied to c-decays directly. 
We worked with the renormalization-scheme-independent Wilson
coefficients. Furthermore, to be consistent, we introduced the effective
Wilson coefficients by taking into account the operator renormalization
to the one-loop order. 

There are some uncertainties in our calculations. While discussing 
direct CP violation, we have to evaluate hadronic matrix
elements where nonperturbative QCD effects are involved. We have worked 
in the factorization approximation, which has not been justified completely
up to now. It has been pointed out that this approximation may
be quite reliable in energetic weak decays\cite{bjorken, dugan}. 
There has also been some discussion on nonfactorizable contributions. In
Ref.\cite{neubert} the authors introduced two phenomenological parameters,
$\epsilon_1$ and $\epsilon_8$, which are scale dependent to parametrize
nonfactorizable effects. The scale dependence of $\epsilon_1$ and 
$\epsilon_8$ cancels that of the Wilson coefficients $c_1$ and $c_2$
and it leads to $a_1^{\rm eff}$ and $a_2^{\rm eff}$. 
In Refs.\cite{chengn, ali},
renormalization-scheme-independent coefficients are used and with the
definition in Eq.(\ref{3dd}) an effective $N_c$ is introduced to describe
nonfactorizable effects. On the other hand, Buras and Silvestrini\cite{buras1}
demonstrated that, in the approach of Ref.\cite{neubert}, 
it is possible to find
a renormalization scheme in which the nonfactorizable parameters 
$\epsilon_1$ and $\epsilon_8$ vanish at any chosen decay scale. In principle,
such a scheme can be determined by experimental data. However, the present
data is not accurate enough. 

We can see from these investigations that more 
work is needed before we can judge the factorization approach. 
Since c-decays are less energetic than b-decays, we expect even more
nonfactorizable effects. In the present work, as in Ref.\cite{chengn, ali},
we introduced an effective value of $N_c$ in Eq.(\ref{3dd}) and assumed 
that it is the same for each $a_i^{\rm eff}\; (i=1,...,6)$. 
The value of $N_c$ should be determined
by experimental data and it will, in general, depend on the 
decay channel, since hadronization
dynamics can be different for each channel. Furthermore, its value
depends on the Wilson coefficients to be used.
We avoid the scheme dependence in Wilson coefficients by using the scheme 
independent ones. However, such coefficients do depend on infra-red regulators
and gauge. In principle, this dependence should be canceled by the
matrix elements of the operators. Furthermore, while discussing the processes
$D^+ \ra \pi^+ \rho^0 (\omega) \ra \pi^+ \pi^+\pi^-$ and 
$D^0 \ra \pi^0 \rho^0 (\omega) \ra \pi^0 \pi^+\pi^-$ we have to evaluate
the matrix elements in some phenomenological quark model.
All these factors may lead to some uncertainty in our 
numerical results. However, as pointed out earlier, since the large CP 
violation we predict is mainly caused by small $t_\rho^f$ in some range
of the phenomenological parameter associated with the breakdown of 
factorization, we expect that our predictions should still provide useful
guidance for future investigations.

\vspace{1cm}

\noindent {\bf Acknowledgment}:
\vspace{2mm}

One of us (Guo) would like to thank A.J. Buras for stimulating communications.
This work was supported in part by the Australian Research Council and
the National Science Foundation of China.


\baselineskip=20pt


\end{document}